# Scattering correction for infrared spectra of biological cells using computational infrared microspectroscopy and deep learning

Sergio G. Rodrigo[1,2*], Ilia L. Rasskazov[3], Luis Martin-Moreno[1,4], Martin Schnell[5,6,7*]

[1] *Instituto de Nanociencia y Materiales de Aragón (INMA), CSIC-Universidad de Zaragoza, Zaragoza, 50009, Spain*

[2] *Departamento de Física Aplicada, Universidad de Zaragoza, Zaragoza, 50009, Spain*

[3] *Independent Researcher, San Jose, California 95124, United States*

[4] *Departamento de Física de la Materia Condensada, Universidad de Zaragoza, Zaragoza, 50009, Spain*

[5] *CIC nanoGUNE BRTA, 20018 Donostia-San Sebastián, Spain*

[6] *Donostia International Physics Center, Donostia-San Sebastian 20018, Spain*

[7] *IKERBASQUE, Basque Foundation for Science, 48013 Bilbao, Spain*

** Corresponding authors. Email: sergut@unizar.es (SGR), schnelloptics@gmail.com (MS)*

**Infrared (IR) microspectroscopy of single biological cells is challenged by strong light scattering, which produces baseline effects and peak distortions in the IR spectra and hinders the direct extraction of chemical information. Current methods for scattering correction typically rely on Mie theory and are accurate only under the assumption that the cell can be approximated by a sphere. Here, we present a framework for the scattering correction of IR absorbance spectra that is based on 3D ellipsoid models and provides efficient scattering correction for both suspended (spherical) and adhered (flattened) cells. Our approach combines deep learning approaches with computational IR microspectroscopy based on the finite-difference time-domain (FDTD) method. The FDTD method generates a synthetic library of realistic training spectra, while the deep learning model enables fast spectral inversion. We demonstrate scattering correction in silico using numerical cell phantoms of cervical cancer cells (HeLa) and show that the true absorption spectra can be inferred from IR absorbance spectra. We further show that the 3D cell dimensions can be recovered from the IR absorbance spectra, highlighting that the inherent light scattering could be exploited to realize the full analytical potential of IR spectroscopy. We anticipate that deep learning-based scattering corrections can be readily extended to increasingly complex sample geometries owing to the flexibility of the FDTD method to model arbitrary geometries.**

## Introduction

Infrared (IR) absorption microspectroscopy is a well-established and robust tool for the label-free analysis of the chemical composition and structural properties of samples, which is used in numerous fields of science and technology [1,2] such as in biomedical [3–5] and forensic applications [6], plant biology [7] and polymer science [8,9]. The commonly utilized mid-IR region spans approximately 4000 - 400 $cm^{-1}$, which corresponds to 2.5 - 25 μm wavelength illumination. Within this spectral region, the fundamental vibrations and associated rotational-vibrational structure in a sample can be probed. While the underlying Beer-Lambert's law of electromagnetic wave attenuation straightforwardly relates the IR absorbance to the optical path length and the absorptivity of the sample, in most of the cases, the recorded signal may be affected by a variety of other mechanisms [10]. Among those mechanisms are scattering and interference phenomena, which represent the case of specific interest since they are especially strong for those samples with spatial structure on length scales comparable to the wavelength of incident electromagnetic radiation. Under these circumstances, light-matter interactions may significantly distort the recorded absorbance spectrum, which complicates the analysis of the recorded data. There are numerous examples of samples where these distortions can occur such as with micron-sized cells [11,12], pollen [13], polymer fibers [14] and beads [15]. On the other hand, the scattering and interference processes encode information on the sample morphology into the absorbance spectrum, which could also be viewed as a useful effect. Particularly, it may be interesting and useful to extract and separate sample morphology and absorbance information from the recorded spectra for obtaining a full analytical insight.

The most important problem to solve in this context is the recovery of the true absorption spectrum from the recorded signal; that is, the recovery of the energy dissipated as heat within the sample (true absorption, inherently free of scattering) from the total loss of light registered by the detector through the optical system (absorbance, contains both absorption and scattering effects). The solution of the inverse problem is quite challenging task due to a variety of spectral distortions emerging from numerous morphologies of samples and setups [16]. Specifically, resonant Mie scattering has been identified as the primary cause for light scattering in case of spherically-shaped samples such as polymer beads or biological cells, creating dispersion-like peak shapes, peak shifts and broad oscillations in the baseline of the spectrum. This finding suggests that light scattering from these type of samples can be described by Mie theory and subsequently removed, yielding a corrected spectra that resembles the true absorption spectra. [12,17–19] The electric field standing wave effect has been identified as a further cause for IR spectral distortions, which describes the reflections at the sample surface and the sample-substrate interference, leading to significant interference (wave patterns) in the IR spectra as we as distortions in peak shape, location and height. [20,21]

To understand the physics underlying these distortions and develop the appropriate strategy for recovering the refractive index of the sample, it is critical to develop theoretical and numerical forward models for electromagnetic light scattering. In this context, seminal works have reported the relevant theory for planar layers [20], grating structures [22], spheres [15], and cylinders [14], clusters of spheres [23,24] and domes [25]. These models can be implemented analytically and allow for fast scattering corrections. Yet, in practice, it is typically required to select the most suitable model based on the specific sample morphology. The sample morphology may vary greatly and thus the accuracy of any scattering correction based on a single model may be limited. This situation is particularly challenging in case of biological cell samples, where cell shapes may vary significantly from nearly spherical in the case of suspended cells to flattened in case of substrate adhered cells. Consequently, it is desirable to have a single model that can provide an accurate description of the light-object interaction for the broadest range of relevant geometries, but such model remains elusive for analytical approaches. Recently, deep learning (DL) has been shown to be powerful and versatile tool in extracting true absorbance spectra from IR spectroscopic data in conjunction with analytical models for electromagnetic scattering, and the potential towards tomographic applications was demonstrated. [26–30] The potential of DL for general spectral preprocessing is also apparent. [31–34] Importantly, deep learning strategies may provide the key to make numerical models practical for IR spectroscopy. The primary advantage of these numerical models lies in their ability to describing the electromagnetic light scattering of arbitrary sample geometries, thus promising a single-model solution for the scattering correction of diverse biological cell samples. Although numerical methods are inherently slow and computationally intensive – rendering them impractical for their standalone use – , deep learning offers a mechanism to accelerate the numerical calculations and potentially enabling fast scattering correction.

In this work, we develop a tandem framework that integrates the finite-difference time-domain (FDTD) method with deep neural networks (DNN). Within this framework, the FDTD method generates realistic IR absorbance and true-absorption spectra to train the DNN, while the DNN facilitates fast scattering correction of IR spectroscopic data by extracting the true absorption spectra from the IR absorbance spectra. Specifically, this approach utilizes 3D ellipsoids as a versatile geometric proxy for a wide range of biological cell morphologies. We demonstrate this approach in silico using numerical cell phantoms of a cervical cancer cell line (HeLa) and show that the true absorption as well as cell size can be accurately extracted of individual IR absorbance spectra.

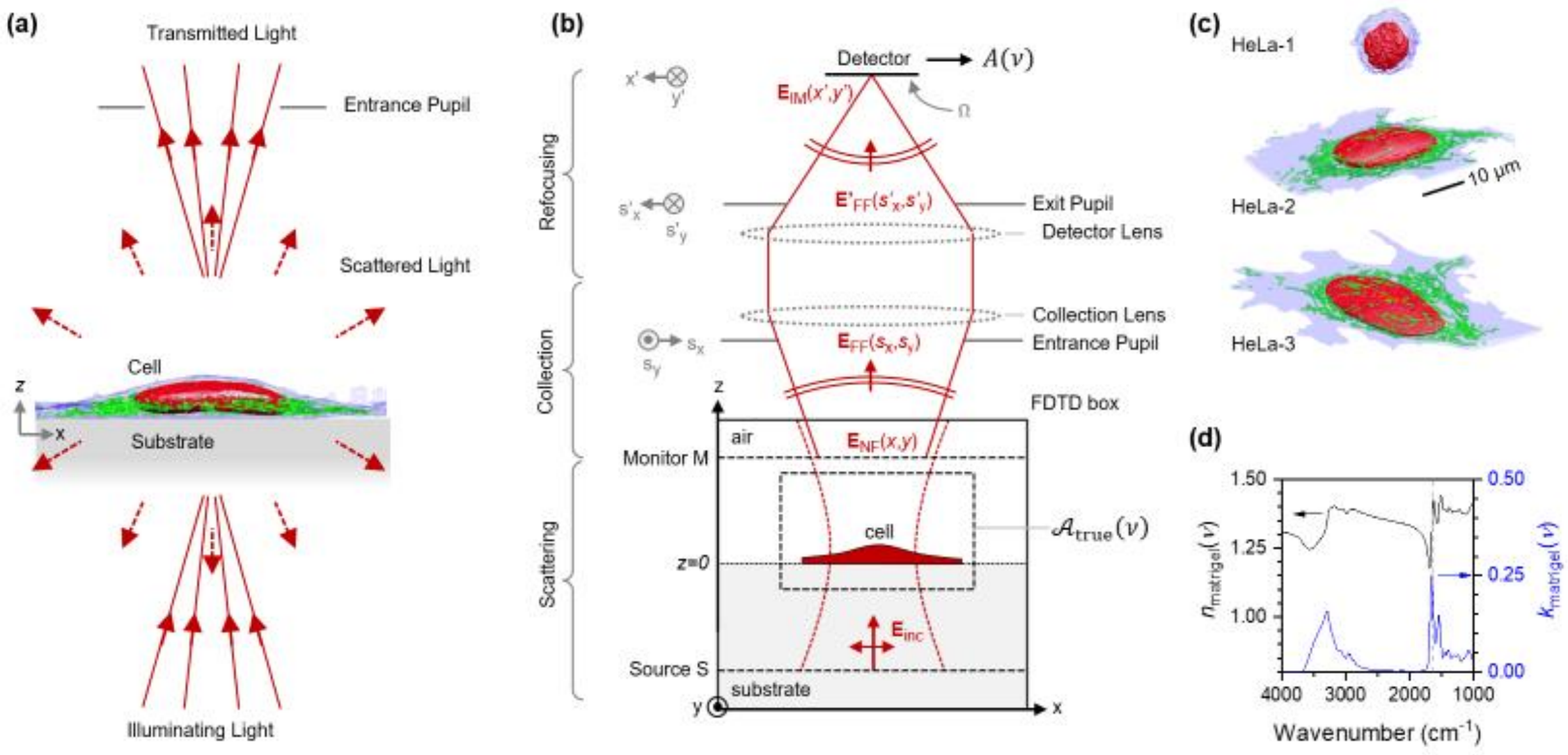

Figure 1. (a) Schematic of IR microspectroscopy performed on a cell adhered to a substrate in transmission. Colors in cell model are blue: cytoplasm, red: cell nucleus, green: mitochondria. (b) Computational IR microspectroscopy system based on FDTD. Symbols see text. (c) Numerical phantoms are constructed from 3D electron microscopy data of cervical cancer cells (HeLa) and (d) the complex refractive index of Matrigel, $\tilde{n}_{\mathrm{matrigel}}(\nu) = n_{\mathrm{matrigel}}(\nu) + ik_{\mathrm{matrigel}}(\nu)$.

## Results

### Computational IR microspectroscopy

Figure 1(a) illustrates the concept of IR microspectroscopy of single biological cells. An IR broadband light beam is focused on the cell. Frequency components that match IR molecular vibrations are absorbed in the cell, which can be detected as a decrease in the intensity of the transmitted beam. The cell is an excellent scatterer for IR light given that the typical dimension of a biological cell is of the order of the wavelength of Mid-IR light. Light scattering outside the entrance pupil leads to an additional loss in the detected intensity of the transmitted beam, i.e. an increase of the apparent absorbance. While light absorption is responsible for the typical molecular line shapes in IR absorbance spectra, light scattering is known for causing a non-resonant baseline in the IR absorbance spectra as well as peak shifts and line shape distortion.

In Fig. 1(b), we describe a method for computational IR microspectroscopy of samples with arbitrary morphology that provides an accurate description of the light-sample interaction and the specifics of light collection of the optical apparatus. We base our work on the numerical method described by Çapoğlu et al. for implementing a virtual imaging system on a computer [35], which was originally applied for generating synthetic microscope images for optical phase-contrast microscopy and later for quantitative phase imaging [36]. Here we adapted this algorithm to IR microspectroscopy. Briefly, the computational IR microspectroscopy system is segmented into scattering, collection and refocusing sections (Fig. 1(b)). First, the light scattering is calculated in the FDTD computational domain as follows. An object sitting on a substrate is illuminated from below with a Gaussian focus by a source S, where the focus is placed on the sample surface ($z = 0$) and a numerical aperture of $NA = 0.6$ is assumed. The object itself may have an arbitrary shape, which is discretized onto the rectangular grid pattern of the FDTD domain, and each point of the grid pattern may be assigned a complex refractive index, $n(\nu) + ik(\nu)$. Light interaction with the object is calculated by solving Maxwell's equation in the time domain using a commercial FDTD solver (Lumerical, Ansys). After the calculation has finished, the field scattered by the object, $\mathbf{E}^{\mathrm{NF}}(x,y,\nu)$, is collected above the object by monitor M, where $\nu$ is the free-space wavenumber of the scattered field. Light collection is implemented by a near-field to far-field transform based on Fourier analysis, producing the field at the far zone, $\mathbf{E}^{\mathrm{FF}}(s_x,s_y,\nu)$, where $s_x, s_y$ are the direction cosines. Plane-wave components $s_x, s_y$ falling outside the numerical aperture of the collecting objective, $\sqrt{s_x^2 + s_y^2} > NA$, are truncated in order to model the entrance pupil of said objective. Note that the objective's numerical aperture is matched to that of the incident Gaussian beam. Light refocusing is implemented by a Fourier transform of the field passing the exit pupil to yield the field distribution at the image plane, $\mathbf{E}^{\mathrm{IM}}(x',y',\nu)$. The signal measured by the detector is then obtained by spatial integration of the intensity of the field over the detector area Ω,

$$I(\nu) = \int_{\Omega} \left|\mathbf{E}^{\mathrm{IM}}(x',y',\nu)\right|^2 \mathrm{d}x'\mathrm{d}y' \ , \qquad (1)$$

We consider the specific implementation of a point mapping system for IR microspectroscopy, where a single large area detector is used for light detection. In this case, the signal measured by the detector is independent of the focusing on the detector and the detector signal can be obtained by spatial integration of the collimated beam exiting the collecting objective [20,22]. That is, it is not needed to integrate over $\mathbf{E}^{\mathrm{IM}}$ as in Eq. (1), instead the detector signal, $I(\nu)$, can already be obtained by integration over $\mathbf{E}^{\mathrm{FF}}$,

$$I(\nu) = \int_{s_x^2+s_y^2 \leq NA} \left|\mathbf{E}^{\mathrm{FF}}(s_x,s_y,\nu)\right|^2 \mathrm{d}s_x \mathrm{d}s_y \ . \qquad (2)$$

Note that the detector signal, $I(\nu)$, directly yields the IR spectrum from the FDTD calculation because the sample's near field, $\mathbf{E}^{\mathrm{NF}}(x,y,\nu)$, and far field distributions, $\mathbf{E}^{\mathrm{FF}}(s_x,s_y,\nu)$, are already spectrally resolved by the FDTD method. This contrasts with FTIR spectroscopy, where the detector signal yields an interferogram and the IR spectrum is obtained by Fourier transform of the interferogram. Nonetheless, the resulting spectral information is equivalent in both cases.

The detector signal, $I(\nu)$, directly yields the unreferenced IR transmission spectrum of the sample. The unreferenced spectra are influenced by instrumental factors such as the transmission function of the beam splitters, the detector responsivity and the spectral characteristics of the light source. These factors can be eliminated by dividing the unreferenced spectrum, $I(\nu)$, with the spectrum from a known reference material, $I_0(\nu)$. This division yields the transmittance, $T(\nu)$, of the sample,

$$T(\nu) = \frac{I(\nu)}{I_0(\nu)} \; . \tag{3}$$

In case of the presented method, the reference spectrum, $I_0(\nu)$, can be obtained straight forwardly by repeating the calculation where the object (cell) is removed. The recorded signal is then expressed as the IR absorbance, $A(\nu)$,

$$A(\nu) = \log_{10} \frac{1}{T(\nu)}. \tag{4}$$

The IR absorbance $A(\nu)$, is the loss of light recorded by the detector (Eq. (4)). The absorbance, $A(\nu)$, describes light scattering and interference effects at the sample, which are relevant for the accurate description for IR spectroscopy of biological cells. In addition, the FDTD method can simultaneously provide the true absorption – also called absorptance –, $\mathcal{A}_{\mathrm{true}}(\nu)$, which is the light energy that is converted into heat due to the excitation of molecular vibrations in the cell. This quantity is free of light scattering and can thus serve as a reference in validating the scattering correction presented below. The true absorption, $\mathcal{A}_{\mathrm{true}}(\nu)$, is obtained by recording the net optical power flow, $P_{\mathrm{abs}}(\nu)$, through a closed volume around the cell (see dashed box labeled $\mathcal{A}_{\mathrm{true}}(\nu)$ in Fig. 1(b)) and dividing it by the power injected by the source S, $P_{\mathrm{in}}(\nu)$,

$$\mathcal{A}_{\mathrm{true}}(\nu) = \frac{P_{\mathrm{abs}}(\nu)}{P_{\mathrm{in}}(\nu)}. \tag{5}$$

For comparison with $A(\nu)$ and $\mathcal{A}_{\mathrm{true}}(\nu)$, it is useful to introduce the bulk absorbance, where scattering effects are ignored and absorption due to molecular vibrations is the only reason for a loss of transmitted light. This bulk absorbance, $A_{\mathrm{bulk}}(\nu)$, is interpreted as being proportional to the concentration of the analytes within the sample by relying on Beer's law,

$$A_{\mathrm{bulk}}(\nu) = \sum_i \alpha_i(\nu) b c_i. \tag{6}$$

where $\alpha_i(\nu) = 4\pi\nu k_i(\nu)$ is the absorptivity of analyte $i$ at wavenumber $\nu$ with imaginary refractive index, $k_i(\nu)$, $b$ is the sample thickness and $c_i$ is the concentration of analyte $i$.

We note that FDTD method is more versatile than analytical models that were developed for IR spectroscopy that take into account light scattering and interference effects for specific sample geometries, including planar layers [20], grating structures [22], spheres [15], and cylinders [14], clusters of spheres [23,24] and domes [25]. Specifically, the presented FDTD method can model the IR absorbance, $A(\nu)$, of objects with arbitrary morphology and with nanoscale spatial resolution in all three dimensions, which is sufficiently accurate to describe the fine details of most biological samples analyzed in IR microspectroscopy.

We further note that it is essential to conduct convergence tests of the presented computational IR microspectroscopy to ensure that the unavoidable numerical error of the FDTD method is sufficiently small. We provide this convergence test in Supplementary Note 1 with the example of a grating structure, for which a semi-analytical model is available from ref. [22]. We find that the FDTD-derived results, $A(\nu)$, show an excellent agreement with the semi-analytical model better than $0.5\%$ (peak error) and $\sim 0.1\%$ (root mean square error).

Having established the FDTD method, we first explore the effect of cellular morphology on the IR absorbance spectra, $A(\nu)$, with the help of numerical cell phantoms (Fig. 1(c)). These cell phantoms are constructed by combining high-resolution morphology information obtained from electron microscopy with refractive index data. Specifically, we use publicly available, high-resolution 3D maps of real HeLa cells from Karabağ et al. (spherical HeLa cells) [37] and from the COSEM initiative of Janelia Farms (flattened HeLa cells) [38]. These 3D maps represent the actual morphology of the cell nucleus (red color in Fig. 1(c)) and the cell membrane (blue color in Fig. 1(c)). Additionally, the 3D maps from the COSEM initiative yield highly detailed maps of the cellular organelles such as mitochondria (green color in Fig. 1(c)). These are not considered further in this Study because they are less relevant for IR microspectroscopy owing to their small size. Nevertheless, such detailed cell maps could find use for building highly-detailed numerical cell phantoms for computational optical microscopy such as quantitative phase imaging, which does have the required spatial resolution to resolve larger cellular organelles [36,39]. As for the refractive index model, we assume that the cells are homogeneous filled by Matrigel, an artificial extracellular matrix, composed of cellular components such as protein and lipids, which resembles typical IR spectra of biological cells and thus often used in IR spectral modelling (Fig. 1(d)). [12,40] This simplification of homogeneously filled cells allows isolation of the effect of cellular morphology on the IR absorbance spectra and investigation of the effects of light scattering and absorption for different cellular morphologies. We consider two types of numerical cell phantoms that can be classified as (i) suspended cells (HeLa-1) with a nearly spherical cell morphology of about 15 µm diameter and (ii) adhered cells (HeLa-2,3) exhibiting a flattened morphology of 3 – 5 µm height and extended lateral dimensions on the scale of 40 – 50 µm. It is expected that Mie resonances are dominant in case of HeLa-1, while HeLa-2,3 cells emphasize thin-film interference effects.

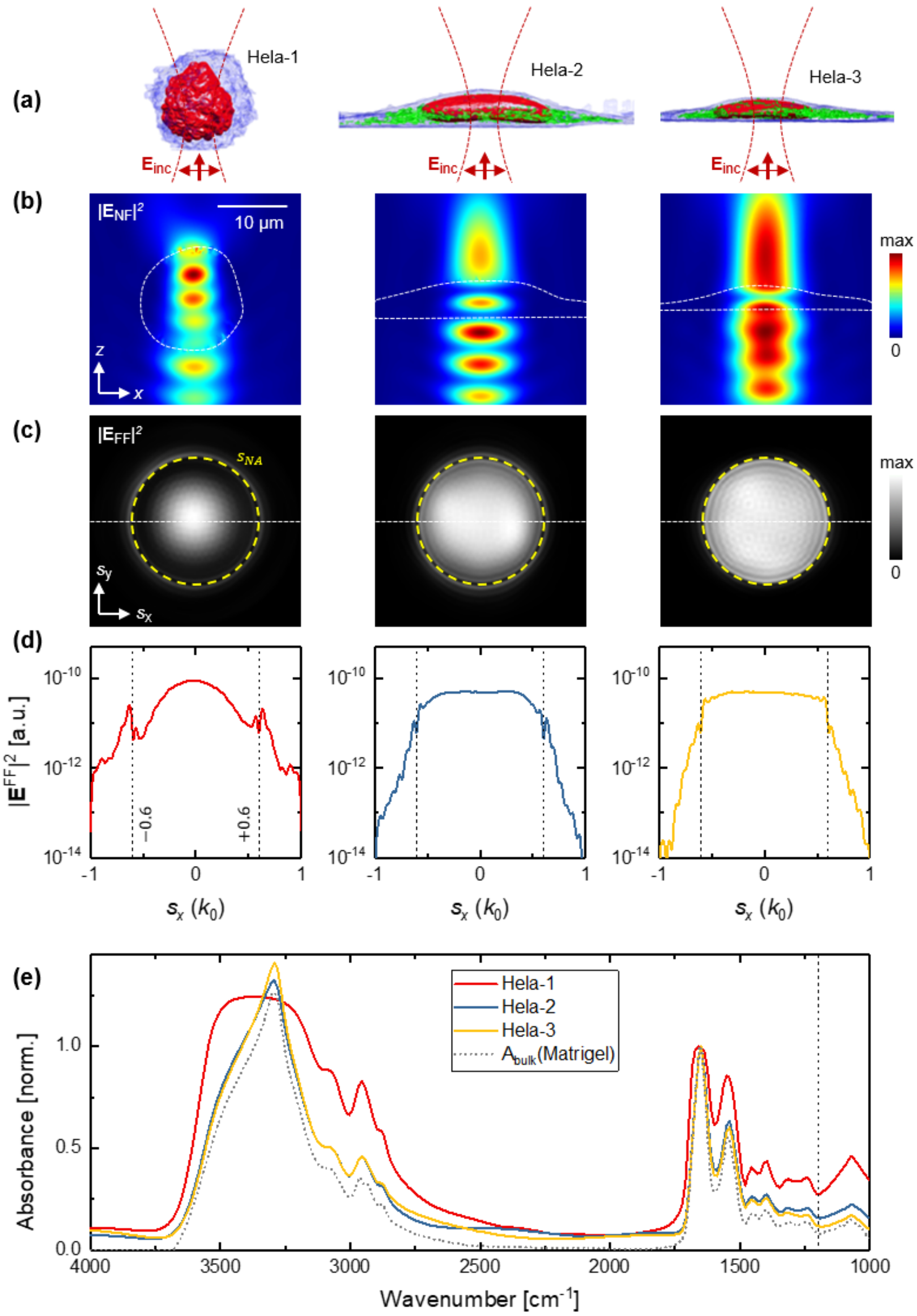

Figure 2: Calculated IR absorbance spectra of numerical cell phantoms. (a) 3D morphology of HeLa cells as obtained from electron microscopy. HeLa-1 was in suspension, resulting in a spherical cell shape, while HeLa-2 and -3 were adhered to substrate, resulting in a flat cell shape. (b) Field amplitude distribution in the vertical (XZ) plane, revealing standing wave interference effects inside of the cells. (c) Transmitted fields as obtained in the far zone, $\mathbf{E}^{\mathrm{FF}}(s_x, s_y, \nu)$. Dashed circle illustrates the collection cone of the objective as defined by its numerical aperture. Data plotted at $\nu = 1200\ \mathrm{cm}^{-1}$ where Matrigel is only weakly absorbing. The different far field scattering characteristics of each cell are apparent. (d) Cross section of $\mathbf{E}^{\mathrm{FF}}(s_x, s_y, \nu)$ for $s_y = 0$ along dashed line in (c). (e) Calculated IR absorbance spectra, $A(\nu)$ (Eq. (4)), normalized to the absorbance at the Amide I band (1655 $\mathrm{cm}^{-1}$) (solid lines). For comparison, the bulk IR absorbance spectra of Matrigel, $A_{\mathrm{bulk}}(\nu)$ (Eq. (6)), is shown (dotted line).

**Scattering effects in biological cells**

To illustrate the effect of cell morphology and light scattering on the IR absorbance spectra, $A(\nu)$, we perform computational IR microspectroscopy for the three HeLa cell phantoms. To this end, it is assumed that the incident beam is focused on the central part of the cell (Fig. 2(a)). Figure 2(b) illustrates the calculated field distribution, revealing the field distribution inside of the cells. In case of the spherical cell (HeLa-1,) the internal field profile is complex and the beam exiting the cell is clearly distorted. The far-field pattern, $\mathbf{E}^{\mathrm{FF}}$, reveals that substantial energy is scattered outside the numerical aperture of the objective (Fig. 2(c), quantified in Fig. 2(d)), which means that light scattering contributes significantly to the IR absorbance spectrum. The calculated IR absorbance spectra reveals strong peak distortion, particularly at higher wavenumbers (3000 to 3500 $cm^{-1}$), and strong baseline effects in the fingerprint region, which are typical for resonant Mie-scattering expected for this round cell shape (red line in Fig. 2(e)). [12] In comparison, the flattened cells (HeLa-2,3) show internal field distributions typical of thin-film interference (Fig. 2(b)), as it can be recognized by comparison with previous studies [20] and with grating structure considered in the Supplementary Figure S2. The corresponding far-field pattern, $\mathbf{E}^{\mathrm{FF}}$, reveals a nearly uniform angular distribution with only weak tails outside of the numerical aperture of the objective, meaning that light scattering contributes only weakly to the IR absorbance spectrum. The resulting IR absorbance spectra, $A(\nu)$, show only small baseline effects and peak distortions (yellow and blue lines in Fig. 2(e)) and overall much higher resemblance with the idealized IR absorbance spectrum of Matrigel, $A_{\mathrm{bulk}}(\nu)$ (dotted line in 2(e)). Our data thus illustrates the strong dependence of the actual cell morphology on IR absorbance spectra, even though all considered cell belongs to the same cell line and are here assumed to be made of the same material (Matrigel). Note that it is critical to consider the optical apparatus – comprising details such as the numerical aperture of the illuminating and collecting lens – in predicting IR spectra as the optical apparatus has a strong effect on the IR spectral shape, which was shown in ref. [15] and illustrated here in Fig. 2. This important detail is of practical relevance because it will increase the accuracy in recovering correct absorption data from the IR absorbance spectra, particularly in the case of spherical cells, as it will be applied later.

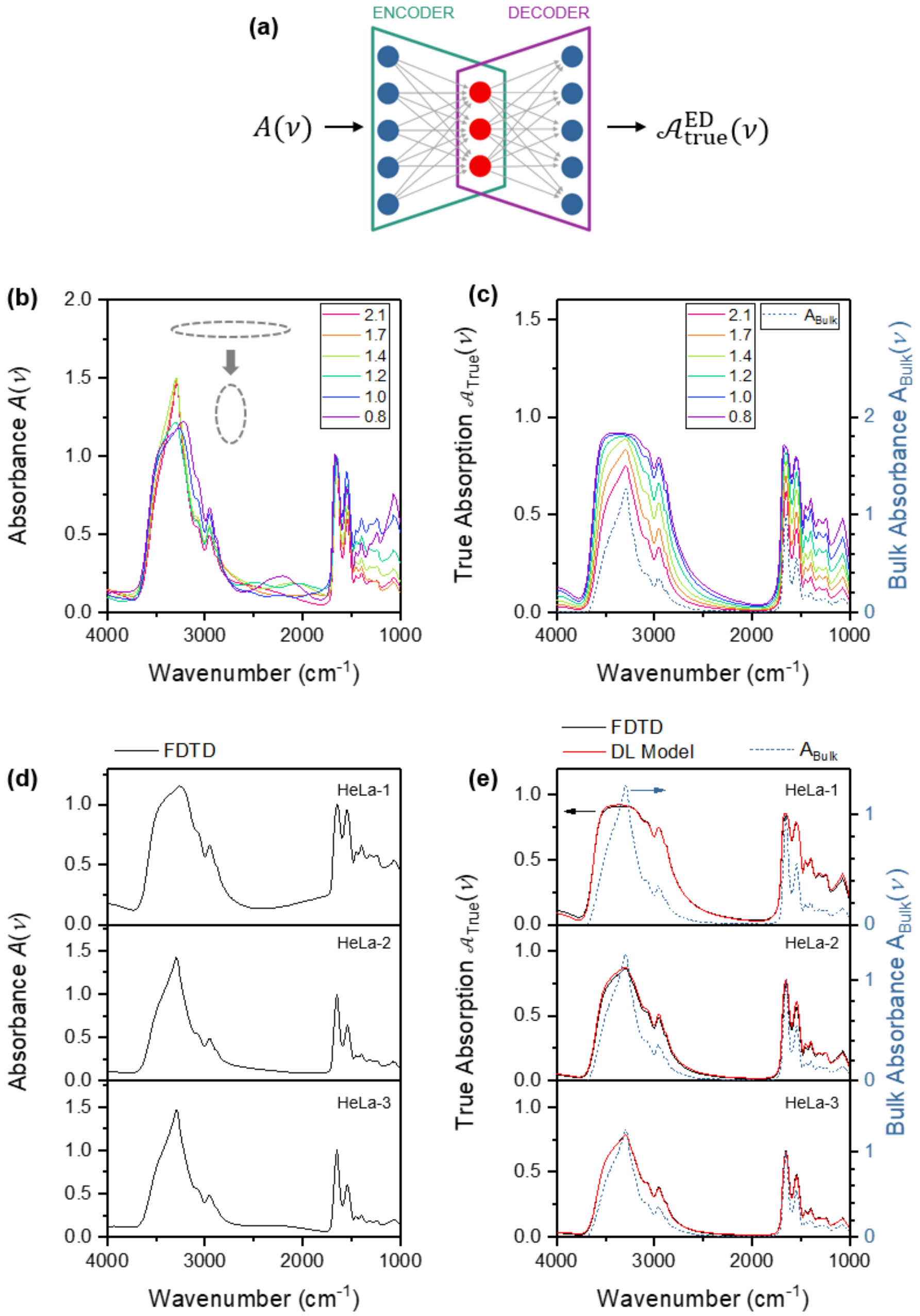

Figure 3: (a) Deep-learning model that accepts the IR absorbance spectrum, $A(\nu)$, from FDTD as input and predicts the true absorption spectrum, $\mathcal{A}_{\mathrm{true}}^{\mathrm{ED}}(\nu)$, as output. (b) Subset of the 3D ellipsoid training data, showing the Absorbance spectra, $A(\nu)$, from FDTD for different volumetric aspect ratios, $\frac{r_x}{R} = \frac{r_y}{R} = 0.8..2.1$, for a constant ellipsoid volume, $V = \frac{4}{3}\pi(5\ \mu\mathrm{m})^3$. (c) The corresponding true absorption spectra, $\mathcal{A}_{\mathrm{true}}(\nu)$, from FDTD. (d) Absorbance spectra, $A(\nu)$, of the nucleus from the three HeLa cells as obtained from FDTD. (e) True absorption spectra, $\mathcal{A}_{\mathrm{true}}^{\mathrm{ED}}(\nu)$, reconstructed from the absorbance spectra, $A(\nu)$, from (d) using the DL model from (a) (solid red line). For comparison, the true absorption spectra, $\mathcal{A}_{\mathrm{true}}(\nu)$, as obtained from FDTD are shown (solid black) as well as the bulk absorbance spectra, $A_{\mathrm{bulk}}(\nu)$, of Matrigel (Eq. (6), dotted black).

**Suppression of IR scattering using the ellipsoid model**

Having established the FDTD method in Figs. 1 and 2, we now present a model for IR microspectroscopy designed to reconstruct the true absorption spectra, $\mathcal{A}_{\mathrm{true}}^{\mathrm{ED}}(\nu)$, directly from IR absorbance spectra, $A(\nu)$, form FDTD. Our approach uses 3D ellipsoids as a geometric proxy for the actual cell morphology, allowing the method to describe both round and flattened cells by design. Traditional physics-based reconstruction would require an iterative optimization loop involving repeated forward FDTD simulations, which is impractical owing to the associated high computational cost. To overcome this drawback, we employ an encoder-decoder-style neural network model (ED) that directly learns the inverse mapping and allows for fast prediction of $\mathcal{A}_{\mathrm{true}}^{\mathrm{ED}}(\nu)$ (Fig. 3(a)), as we will show next.

We first apply the FDTD method from Fig. 1(b) to generate a large set of training data for the ED model, where we consider 3D ellipsoids of different volumes, volumetric aspect ratios and material densities (see Methods). Briefly, we vary the three semi-axis dimensions of the ellipsoid, $r_x$, $r_y$ and $r_z$, as well as the rescale the imaginary refractive index of Matrigel, $k(\nu)$, by setting the Amide I peak height to different values, $k_m$. This four-dimensional parameter space yields a diverse range of geometries, spanning from prolate (elongated in $z$) to spherical to highly oblate (very flat in $z$) forms, and molecular vibrations of different strenght. Thus, this single training set provides a comprehensive morphological description of both the spherical and flattened biological cells. The FDTD method yields both the IR absorbance spectra, $A(\nu)$ (Eq. (4)), and the true absorption spectra, $\mathcal{A}_{\mathrm{true}}(\nu)$ (Eq. (5)), for each configuration, yielding a total of 11,000 training spectra. Figure 3(b,c) illustrates a small subset of this training data set, revealing the transition from Mie scattering to thin-film interference effects as the volumetric aspect ratio is scaled from the prolate to ($r_x/R = r_y/R = 0.5$) to oblate ellipsoids ($r_x/R = r_y/R = 2.1$).

Having trained the DL model, we proceed to validate the accuracy of the DL model in silico. To this end, we predict the absorption spectra, $\mathcal{A}_{\mathrm{true}}^{\mathrm{ED}}(\nu)$, from the calculated absorbance spectra, $A(\nu)$, using the DL model and compare it to the true absorption, $\mathcal{A}_{\mathrm{true}}(\nu)$, as obtained with FDTD. As a first step, only the cell nucleus of the three HeLa cells is considered (red regions in Fig. 1(c)). This is motivated by the fact that the nucleus is largest component of the cell and is widely recognized in IR spectroscopy as the primary origin of Mie scattering in single biological cells. Figure 3(d) shows the absorbance spectra of the cell nuclei, $A(\nu)$, as obtained from FDTD model (black solid lines). Figure 3(e) displays the predicted true absorption spectra, $\mathcal{A}_{\mathrm{true}}(\nu)$, obtained with the DL model (red solid lines), demonstrating very good agreement with the FDTD-derived true absorption spectra, $\mathcal{A}_{\mathrm{true}}(\nu)$ (black lines ). The root-mean-square error, $\Delta\mathcal{A}_{\mathrm{true}} = \sqrt{\sum_i \left(\mathcal{A}_{\mathrm{true}}^{\mathrm{ED}}(\nu_i) - \mathcal{A}_{\mathrm{true}}(\nu_i)\right)^2}$, quantifies the error to less than $\Delta\mathcal{A}_{\mathrm{true}} < 0.018$ for all three cells.

Figure 3(e) shows the benefit of reconstructing the true absorption spectra, $\mathcal{A}_{\mathrm{true}}^{\mathrm{ED}}(\nu)$, with the presented DL model. The effect of Mie scattering and thin-film interferences is visibly reduced in $\mathcal{A}_{\mathrm{true}}(\nu)$, compared to $A(\nu)$ in case of the cells (cf. Fig 3(d) and 3(e)) and the ellipsoids (cf. Fig. 3(b) and 3(c)). Particularly, the silent spectral region of 2000 to 2500 $\mathrm{cm}^{-1}$ is largely smooth and flat, as expected, while the fingerprint spectral region of $1000$ and $1500\ \mathrm{cm}^{-1}$ shows a more consistent peak structure, which facilitates biological interpretation by using $\mathcal{A}_{\mathrm{true}}(\nu)$ (Fig. 3(c,e)) instead of $A(\nu)$ (Fig. 3(b,d)). Note that peak shape distortions are still observed in $\mathcal{A}_{\mathrm{true}}(\nu)$. This is expected because $\mathcal{A}_{\mathrm{true}}(\nu)$ is modulated by the internal field distribution of the ellipsoids and cells and thus different to the bulk IR absorbance, $A_{\mathrm{bulk}}(\nu)$, where it is essentially assumed that a plane wave is passing through a homogenous medium. Importantly, the presented FDTD method can visualize this internal field distribution in actual cell morphologies (Fig. 2(b)), which may help to improve our understanding of how cell morphology changes the true absorption spectra. The ultimate solution to the scattering problem is the reconstruction of the refractive index, yielding directly the strength and frequency of the molecular vibrations. However, this effort would shift the task from qualitative signal processing (as presented here) to quantitative measurement of material properties, which entails much more stringent validation requirements for the presented FDTD method and the DL model.

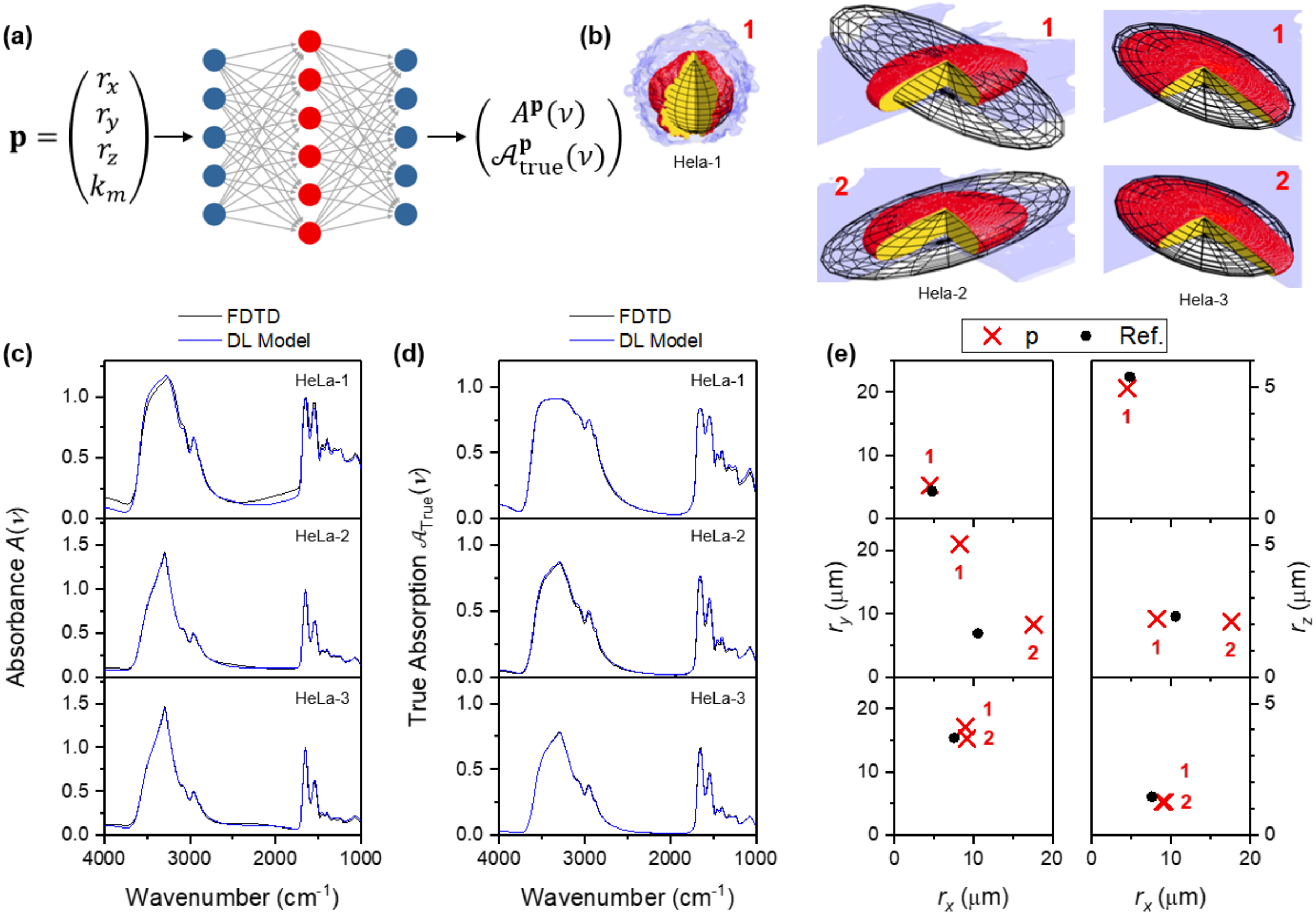

Figure 4: (a) Deep-learning model that accepts the model parameter, $\mathbf{p} = (r_x, r_y, r_z, k)$, as input and predicts either the absorbance spectrum, $A^{\mathbf{p}}(\nu)$, or the true absorption spectrum, $\mathcal{A}^{\mathbf{p}}_{\text{true}}(\nu)$, as output. (b) Extraction of ellipsoid size by inversion of DL model in (a) using the absorbance spectra, $A(\nu)$, of cell nuclei from Fig. 3(d). The first two best matches for $\mathbf{p}$ are plotted as an ellipsoid on top of the real morphology of the cell nuclei. (c) The corresponding absorbance and true absorption spectra, $A^{\mathbf{p}}(\nu)$ and $\mathcal{A}^{\mathbf{p}}_{\text{true}}(\nu)$, for the best match for $\mathbf{p}$ (blue lines) compared to the cell nuclei spectra obtained from FDTD (black lines). (e) The first two matches for $\mathbf{p}$ (red crosses) compared to the dimensions as obtained by fitting an ellipsoid directly to the cell nuclei morphology (black dots).

**Inference of sample morphology from IR spectra**

So far, the encoder-decoder-style (ED) model from Fig. 3 is independent by design of any explicit reference to morphology parameters as it simply predicts $\mathcal{A}_{\text{true}}(\nu)$ from $A(\nu)$. To establish a clear connection between sample morphology and absorbance, $A(\nu)$, we present two additional DL models in Fig. 4(a) that are specifically tailored to the ellipsoid geometry used in the training data. These DL models act as surrogate models to accelerate the computationally intensive FDTD calculations, thereby enabling rapid multi-parameter optimization loops to achieve fast signal inversion. Specifically, these DL models take the semi-axes, $r_x$, $r_y$ and $r_z$, as well as the material density parameter, $k$, as input parameter $\mathbf{p}$ and predict either $A^{\mathbf{p}}(\nu)$ or $\mathcal{A}^{\mathbf{p}}_{\text{true}}(\nu)$. We apply these DL models to invert the absorbance and true absorption spectra from the cell nuclei, $A(\nu)$ and $\mathcal{A}_{\mathrm{true}}(\nu)$, from Fig. 3(d,e) to obtain $\mathbf{p}$. This inversion essentially fits an ellipsoid to the cell nuclei based on $A(\nu)$ and $\mathcal{A}_{\mathrm{true}}(\nu)$. Briefly, we perform multiple inversion runs with different seed values using the gradient descent method and perform k-means segmentation of the results to determine the centroids. The centroids are the results of the fitting, ordered by the loss function (see Methods for details). Figure 4(b) illustrates the fitted ellipsoids for the first (1) and second (2) best fit as a mesh grid (black lines) around the 3D morphology of the cell nuclei (red color). Figure 4(e) further quantifies the two best fit results $\mathbf{p}$ (red crosses) compared to an ellipsoid fit directly to the cell nuclei morphology (black dots, see Methods). Table 2 lists all fit results.

Overall, we obtain a good to very good agreement for all cells. Specifically, the semi-axis, $r_z$, and the material density parameter, $k_m$, were correctly determined, which are the two critical parameter in determining the IR absorbance spectra. The lateral dimensions are correctly fitted in case of the cells HeLa-1 and HeLa-3. On the contrary, the results of HeLa-2 show some ambiguity, which warrants further discussion. First, there exist certain symmetries in the DL model $\mathbf{p} \rightarrow \left(A^{\mathbf{p}}(\nu), \mathcal{A}_{\mathrm{true}}^{\mathbf{p}}(\nu)\right)$ when the cells are elongated in *x* or *y*, as it is the case with HeLa-2. The incident beam of the computation IR microspectrometer is polarized, but this polarization provides only for a weak discriminator for detecting the orientation of an elongated cell or ellipsoid. Second, the cell nucleus of HeLa-2 is curving slightly upwards creating a small bulge, which deviates from the ellipsoid form and thus shows the limitation of the ellipsoid training data set. Extending the training data set by more complex geometries beyond ellipsoids could be a solution to improving fit accuracy.

To verify that the retrieved values for $\mathbf{p}$ are correct, we apply the DL models in Fig. 4(a) in the forward direction to obtain the corresponding absorbance and true absorption spectra, $A^{\mathbf{p}}(\nu)$ and $\mathcal{A}_{\mathrm{true}}^{\mathbf{p}}(\nu)$ (blue lines in Fig. 4(c,d)). As visual inspection reveals, we obtain an excellent agreement with the calculated absorbance and true absorption spectra of the cell nuclei, $A(\nu)$ and $\mathcal{A}_{\mathrm{true}}(\nu)$, obtained from FDTD (black lines). This finding establishes a clear connection between scattering correction (Fig. 3) and morphology retrieval (Fig. 4).

Note that the retrieval of the morphology parameters, $\mathbf{p}$**,** is done using a multi-objective inversion approach and thus requires knowledge of both $A(\nu)$ and $\mathcal{A}_{\mathrm{true}}(\nu)$. While the absorbance $A(\nu)$ is experimentally easily accessible and specifically the quantity measured in many IR microspectrometers, experimental measurement of the true absorption, $\mathcal{A}_{\mathrm{true}}(\nu)$, requires local temperature measurements. The latter are possible with photothermal methods in principle, but they are generally difficult to do. Therefore, the presented inversion approach in Fig. 4 is designed in a way to first predict $\mathcal{A}_{\mathrm{true}}^{\mathrm{ED}}(\nu)$ from $A(\nu)$ using the autoencoder-based model from Fig. 3(a), and then to perform the retrieval of $\mathbf{p}$, yielding the results from Fig. 4.

|  | Fit Type | $r_x$ | $r_y$ | $r_z$ | $k_m$ |
|---|---|---|---|---|---|
| HeLa-1 | Morphology | 4.81 | 4.32 | 5.37 | 0.250 |
|  | Deep Learning #1 | 4.44 | 5.23 | 4.94 | 0.272 |
| HeLa-2 | Morphology | 10.56 | 6.89 | 2.30 | 0.250 |
|  | Deep Learning #1 | 8.26 | 21.00 | 2.20 | 0.233 |
|  | Deep Learning #2 | 17.63 | 8.29 | 2.09 | 0.242 |
| HeLa-3 | Morphology | 7.60 | 15.35 | 1.44 | 0.250 |
|  | Deep Learning #1 | 8.93 | 17.13 | 1.26 | 0.257 |
|  | Deep Learning #2 | 9.18 | 15.24 | 1.25 | 0.260 |

Table 1: Cell dimensions as obtained by inversion of the deep learning model in Fig. 4, the best two results are shown. For comparison, a 3D ellipsoid fit to the morphology of the cell nucleus is given.

**Discussion and Conclusion**

We have developed a computational IR microspectroscopy system that can provide realistic IR absorbance spectra for the training of deep learning models. Importantly, the computational data align closely with experimental data because the optical apparatus is taken into account [15,22,41]. Further, the IR absorbance spectra are free of instrumental imperfections such as lens aberrations or drift. Therefore, the presented method is a promising way for the fast generation of large amounts of high-quality training data that can be used for deep learning models for IR microspectroscopy, thus potentially obviating careful experimental studies that are otherwise needed to generate such data.

We further have demonstrated the utility of numerical cell phantoms for understanding the signal formation in IR microspectroscopy. These cell phantoms allow for an objective comparison between various microscopy configuration or methods [36], and in this Study, they were used to validate the scattering correction provided by the presented deep-learning models. We anticipate that both the quantity and level of detail of numerical cell phantoms will increase. 3D electron microscopy is already a well-established source for highly detailed cell and tissue morphology data set, and the amount of publicly made available data sets is increasing thanks to open-access efforts. [38] On the other hand, measurement of the IR refractive index inside of cells remains a challenge. However, emerging techniques such as near-field microscopy have shown considerable promise, [42] and these research efforts should be intensified to create more accurate numerical cell phantoms models.

The presented DL-based scattering correction assumes prior knowledge about the sample material. Matrigel was chosen as the cell material, and only the strength of the molecular vibrations was varied in the training data set using the scaling parameter, $k_m$, to introduce this variation in the training data set. This choice was made on purpose to study and isolate the effect of sample morphology (light scattering) on the IR absorbance spectra. We anticipate that the presented DL models can be extended for real world applications, where the sample material is not known beforehand. To this end, the DL model in Fig. 3 ($A(\nu) \rightarrow \mathcal{A}_{\mathrm{true}}(\nu)$) – which is independent of any specific geometry – could be expanded by considering an ellipsoid training data set where a variety of materials is assumed.

## Methods

### Numerical cell phantoms

The numerical cell phantoms were constructed as follows. The morphology data was downloaded from the cited citations. For example, in case of COSEM, the AWS CLI tool was used to download the data sets from the link provided on the openorganelle page at Janelia Farm (https://openorganelle.janelia.org/). In case of HeLa-2, the link that was used is s3://janelia-cosem-datasets/jrc_hela-2/. We used the "s2" version of the data set, meaning that the pixel size was 16 nm in the *x*, *z*-directions and 20.96 nm in the *y*-direction. The data was converted from the N5 format to image format (.tiff) using FIJI [43]. The cell nucleus was directly provided as a segmentation layer. To obtain the entire cell shape, we applied a fill operation to the plasma membrane segmentation layer in matlab. Both data sets were downsampled to 48 nm in *x*,*y*,*z*-directions to save memory and make them compatible with the square grid of the FDTD simulations. The so-obtained data was exported as a text file so they can be important as a binary structure into Lumerical FDTD.

The pure absorbance template spectrum of Matrigel was obtained from reference [44] (https://github.com/GardnerLabUoM/RMieS/archive/master.zip), scaled to set the maximum height of the Amide I peak to $k(1655 \text{ cm}^{-1}) = 0.25$ and then assigned to the imaginary part of the refractive index, $k_{\text{matrigel}}(\nu)$ as in ref. [45]. The corresponding real part of the refractive index, $n_{\text{matrigel}}(\nu)$, was then calculated with Kramers-Kronig using a background refractive index of $n_{\text{inf}} = 1.34$. The so-obtained complex-valued refractive index data (Fig. 1(d)) was imported into FDTD. To ensure a good fit of all relevant molecular resonances of Matrigel in the Mid-IR range, the number of fit coefficients was set to 35, a passive fit was used and the option of "improve stability" was set to No. The cell morphology data was then imported into FDTD as a 3D binary import, choosing the correct pixel size of 48 nm and choosing Matrigel as the material.

### FDTD calculations

The FDTD calculations were done with Lumerical (Ansys) using Version XXXX. The FDTD simulation was configured as follows. The width of the FDTD box was set to 100 µm (in x and y) and the height was set to 20 µm (z), and 24 PML layers are chosen at all six boundaries. The simulation time of the FDTD box was set to 6000 fs to ensure that the resonances of the Matrigel are properly captured. A global mesh setting of 192 nm (in xyz) was used. A Gaussian source was placed at $z = -3.456$ µm injecting the fields backwards (in +z) with polarization in x-direction. The bandwidth of the source was set to span 2.5 µm to 10 µm (4000 to 1000 $\text{cm}^{-1}$) to cover the Mid-IR spectral range. To ensure an achromatic focusing across this range, we activated the frequency dependent profile function of Lumerical and set the number of frequency profiles to 60. A 2D field profile monitor was set at $z = +3.456$ μm to capture the transmitted light. Spatial downsampling (factor 4) was used to save memory. A mesh override region was defined around the numerical cell phantom to ensure proper sampling of the cell morphology, covering 30 μm width (xy) and 7.2 μm height (z) with a mesh refinement setting to 96 nm. A set of two simulations is run for each cell: A simulation with the cell present ("sim") and a second simulation with the cell removed ("ref"), the latter serving as a reference simulation for signal referencing, as explained in the main text. For each simulation, a near-field-to-far-field transform was performed and the absorbance spectra, $A(\nu)$, are calculated following the steps outlined in the main text. In parallel, a transmission monitor box is defined around the sample, which measured the net optical power flow through the box, $P_{\text{abs}}(\nu)$. This data is divided by the power injected by the source, $P_{\text{in}}(\nu)$, to obtained the true absorption spectrum, $\mathcal{A}_{\text{true}}(\nu)$. For simplicity and clarity, in this paper we ignore possible effects of the substrate on the IR absorbance spectra, $A(\nu)$, by removing the substrate from the FDTD calculation, i.e. $n_{\text{Substrate}} = 1$ in Fig. 1(b).

The FDTD calculations were also applied to generate the training data sets consisting of ellipsoids as shown in Fig. 3. To this end, we replace the numerical cell phantom with a 3D ellipsoid and vary volume, aspect ratio and material density. The ellipsoid volume, $V = \frac{4}{3}\pi R^3$ was varied in the range from $R =$ 2..7 μm in steps of 0.5 μm. The volumetric aspect ratios of the ellipsoid semi-axes, $r_y/R$ and $r_x/R$, were varied independently for values 0.8, 0.97, 1.17, 1.41, 1.71, 2.06, 2.49, 3. The remaining semi-axis was adjusted with $r_z = R^3/r_x r_y$.. Finally, the strength of the molecular vibrations of Matrigel were scaled by setting the height of the Amide I peak (1655 $\text{cm}^{-1}$), $k_m$, to different values ranging from $k_m = 0.05\,..\,0.50$ in steps of 0.05,

$$k(\nu) = k_{\text{matrigel}}(\nu) \cdot \frac{k_m}{k_{\text{matrigel}}(1655\ \text{cm}^{-1})}, \tag{7}$$

where $k_{\text{matrigel}}(\nu)$ is the original Matrigel refractive index and $k(\nu)$ the scaled imaginary refractive index. For each variation, the corresponding real part of the refractive index, $n(\nu)$, was obtained with Kramers-Kronig and using a background refractive index of $n_{\text{inf}} = 1.34$. The training data set therefore contains four independent variables and a total of 11,000 pairs of Absorbance spectra, $A(\nu)$, and true absorption spectra, $\mathcal{A}_{\text{true}}(\nu)$, were generated.

**Deep-learning model (Fig. 3)**

A fully connected encoder-decoder neural network was implemented using Keras with a TensorFlow backend to predict true absorption (heat dissipation) spectra, $\mathcal{A}_{\text{true}}(\nu)$, from absorbance spectra, $A(\nu)$. Both the input and output dimensions corresponded to the number of spectral wavelengths (251). The network architecture consisted of an encoder–decoder structure with three encoding layers and three decoding layers. The encoder compressed the input spectra through successive dense layers containing 50 and 5 neurons with ReLU activation functions, followed by a latent-space representation of 4 neurons. The decoder reconstructed the target absorption spectra through symmetric dense layers of 5 and 50 neurons, ending in a linear output layer matching the spectral dimensionality. Glorot normal and Glorot uniform initializations were used for the output layer weights and biases, respectively.

The training dataset comprised 11,000 spectra, of which 10% were reserved for validation. The model was trained using the Adam optimizer and mean squared error (MSE) loss function for 1000 epochs with a batch size of 32. Early stopping with restoration of the best weights and model checkpointing were employed during training to prevent overfitting and preserve the best-performing model. During training, validation losses as low as $1.64 \times 10^{-4}$ were achieved, demonstrating high reconstruction accuracy between absorbance and absorption spectra.

**Deep-learning model (Fig. 4)**

A fully connected feed-forward neural network was implemented using the Keras library with a TensorFlow backend as a surrogate model for finite-difference time-domain (FDTD) simulations. The network input consisted of four parameters, $\mathbf{p} = (r_x, r_y, r_z, k_m)$, corresponding to the three ellipsoid semi-axes and the $k_m$ parameter for scaling the imaginary refractive index of Matrigel, $k(\nu)$. The output layer contained one neuron per spectral wavelength value to be predicted. The predicted spectra corresponded either to absorbance, $A(\nu)$ or true absorption (heat dissipation), $\mathcal{A}_{\text{true}}(\nu)$. The architecture comprised four hidden layers, each with 64 neurons and sigmoid activation functions, followed by batch normalization layers to improve training stability. The output layer used a linear activation function to enable continuous-valued spectral predictions.

The model was trained using the RMSprop optimizer with a learning rate of 0.01 and momentum of 0.001. Mean squared error (MSE) was employed both as the loss function and evaluation metric. Training was performed for 1000 epochs with a batch size of 64, using 20% of the spectra as validation data to monitor performance. The best-performing model weights were automatically saved during training through a model checkpoint callback. During training, MSE values as low as $1.13 \times 10^{-4}$ were achieved, indicating high predictive accuracy of the surrogate model with respect to the original FDTD-generated spectra.

**Multi-objective Model (Fig. 4)**

An inverse design optimization procedure was implemented using two pretrained surrogate models and an auxiliary encoder-decoder model within the Keras framework running on a TensorFlow backend. The goal of the optimization was to retrieve physically meaningful structural (morphology) parameters from the absorbance spectrum, $A(\nu)$, obtained from FDTD while ensuring consistency with both absorbance and true absorption (heat dissipation) predictions, $A(\nu)$ and $\mathcal{A}_{\text{true}}(\nu)$, respectively. The target spectrum corresponded to experimentally accessible absorbance data, $A(\nu)$, whereas the corresponding true absorption spectra, $\mathcal{A}_{\text{true}}(\nu)$, is difficult to measure experimentally (local temperature measurement would be required) and was thus generated through a learned mapping with the ED model.

In detail, the optimization relied on two surrogate models: the model $\mathbf{p} \to A^{\mathbf{p}}(\nu)$ predicts the absorbance spectra from the design parameters, and the model $\mathbf{p} \to \mathcal{A}_{\text{true}}^{\mathbf{p}}(\nu)$ predicts the true absorption spectra (Fig. 4(a)). In addition, an encoder-decoder-style model, $A(\nu) \to \mathcal{A}_{\text{true}}^{\text{ED}}(\nu)$, was used to infer a physically plausible true absorption spectrum conditioned on the target absorbance (Fig. 3(a)). The combined loss function, $\mathcal{L}(\mathbf{p}) = \gamma \sum_i |A^{\mathbf{p}}(\nu_i) - A(\nu_i)|^2 + \beta \sum_i \left|\mathcal{A}_{\text{true}}^{\boldsymbol{p}}(\nu_i) - \mathcal{A}_{\text{true}}^{\text{ED}}(\nu_i)\right|^2$, was defined as a weighted sum of two mean squared error (MSE) terms: one enforcing agreement between predicted and target absorbance spectra, and another enforcing consistency between the predicted absorption spectrum and the autoencoder-generated absorption representation.

Gradient-based optimization was performed directly in the design-parameter space using automatic differentiation with a learning rate of $10^{-3}$. The parameters were constrained within a normalized range $[0, 1]$using clipping after each update step. The procedure was initialized from a perturbed seed design and repeated over multiple independent runs (M = 50) to improve robustness against local minima. Each run consisted of 10,000 optimization steps using the Adam optimizer. Crucially, evaluating this total of 500,000 design iterations would be unfeasible using the FDTD simulations. Using the DL models, optimization time was reduced to approximately 20 minutes. Furthermore, the encoder-decoder model introduces a direct inverse-mapping capability that has no classical counterpart. Loss convergence was monitored throughout the optimization process, with logarithmic decay behavior typically observed, and final optimized parameter sets were recorded together with their corresponding minimum loss values. We repeated the optimization process by means of the differential evolution algorithm finding consistent results.

**Acknowledgement**

This work was financially supported by the Spanish "Ministerio de Ciencia, Innovación y Universidades", MICIU/AEI/ 10.13039/501100011033, through grants PID2023-148359NB-C21, PID2023-148359NB-C22 (also funded by ERDF/EU), RYC2023-044188-I (also funded by ESF+), CEX2018-000867-S (DIPC), CEX2023-001286-S (INMA) and CEX2020-001038-M (NANOGUNE). We also acknowledge the Aragón Regional Government through project QMAD (E09_23R). Funded by the IKUR Strategy on behalf of the Basque Government's Department of Science, Universities and Innovation. The authors thank Rohit Bhargava (UIUC, USA) for providing the source code for the semi-analytical model from ref. [46] that was used for the validation of the FDTD results in this work.

**Author contributions.**

M.S. and I. R. developed and validated the FDTD model for computational IR microspectroscopy. S.G. R. and L.M.M. developed and validated the deep learning models for the scattering correction. S.G.R., L.M.M. and M.S. performed the data analysis. S.G.R., L.M.M. and M.S. wrote the manuscript with input from all authors. All authors contributed to scientific discussions.

**Competing interests**

The authors declare no competing interests.

**Supporting Information Available**

Supplementary Note 1. Convergence of the FDTD calculations

**Supplementary Information to**

**Scattering correction for infrared spectra of biological cells using computational infrared microspectroscopy and deep learning**

Sergio G. Rodrigo[1,2*], Ilia L. Rasskazov[3], Luis Martin-Moreno[1,4], Martin Schnell[5,6,7*]

[1] *Instituto de Nanociencia y Materiales de Aragón (INMA), CSIC-Universidad de Zaragoza, Zaragoza, 50009, Spain*

[2] *Departamento de Física Aplicada, Universidad de Zaragoza, Zaragoza, 50009, Spain*

[3] *Independent Researcher, San Jose, California 95124, United States*

[4] *Departamento de Física de la Materia Condensada, Universidad de Zaragoza, Zaragoza, 50009, Spain*

[5] *CIC nanoGUNE BRTA, 20018 Donostia-San Sebastián, Spain*

[6] *Donostia International Physics Center, Donostia-San Sebastian 20018, Spain*

[7] *IKERBASQUE, Basque Foundation for Science, 48013 Bilbao, Spain*

** Corresponding authors. Email: sergut@unizar.es (SGR), schnelloptics@gmail.com (MS)*

| Symbol | Meaning |
| --- | --- |
| $A(\nu)$ | The absorbance spectrum as obtained from the FDTD calculation by registering the loss of light at the detector of the optical apparatus. |
| $\mathcal{A}_{\mathrm{true}}(\nu)$ | The true absorption spectrum – also called absorptance – as obtained from the FDTD calculation, where the net optical power flow through a closed box around the cell is measured. |
| $A_{\mathrm{bulk}}(\nu)$ | The bulk absorbance as obtained by Beer's law. |
| $\mathcal{A}_{\mathrm{true}}^{\mathrm{ED}}(\nu)$ | The true absorption predicted from the absorbance $A(\nu)$ using the encoder-decoder-style neural network model in Fig. 3 of the main text. |
| $\mathbf{p}$ | Parameter $\mathbf{p} = \left(r_x, r_y, r_z, k\right)$ for the DL model in Fig. 4 of the main text which describes the semi-axes of the ellipsoid, $r_x, r_y$ and $r_z$, and the scaling parameter, $k_m$, for the imaginary refractive index of Matrigel, $k(\nu)$. |
| $A^{\mathbf{p}}(\nu)$ | The absorbance spectra predicted based on parameter $\mathbf{p}$ using the DL model in Fig. 4 of the main text. |
| $\mathcal{A}_{\mathrm{true}}^{\mathbf{p}}(\nu)$ | The true absorption spectra predicted based on parameter $\mathbf{p}$ using the DL model in Fig. 4 of the main text. |
| $k_{\mathrm{matrigel}}(\nu)$ | The original imaginary refractive index for Matrigel, as obtained from ref. [1]. These data are used for the numerical cell phantoms. |
| $k(\nu)$ | The scaled imaginary refractive index for Matrigel by setting the peak height of the Amide I band to $k_m$. These data are used for the ellipsoid training data set. |

Table S1: Glossary of notation.

**Supplementary Note 1: Convergence of the FDTD calculations**

Numerical FDTD calculations are the basis for the *computational IR microspectrometer* method that is presented in main text. In the following, we present a study to understand sources of numerical error and provide steps that can be taken to reduce that error to an acceptable level. We will show that the *computational IR microspectrometer* method has specific sources of error such as the width of the FDTD box, the global mesh setting and the number of field profiles of the source. We will provide an absolute convergence test by comparison with a semi-analytical model of a grating structure and show that the numerical error in the absorbance spectra, $A(\nu)$, can be reduced to the order of 0.1%.

***Configuration of the FDTD simulation***

Figure S1(a) shows the schematic of the *computational IR microspectrometer* method for the case where the sample is a grating structure. The grating structure can fulfill multiple roles for testing the convergence of the FDTD calculations. First, if the focus is placed on center of the slab, it behaves similar to a homogeneous thin film and we can test for distortions in the IR transmission spectra arising from the sample-substrate structure. [2] Second, if the focus is placed near the edge of the slab, we can test for distortions arising from the scattering of the incident beam at the edges. [3] The resulting transmitted beam shows a complex scattering pattern, where the energy of the incident beam is redistributed between inside and outside the aperture of the collecting lens that would otherwise be fully incident on the detector. This effect – the reduction of the IR transmission signal through light scattering – provides opportunity to validate that the optical apparatus is accurately implemented. To this end, the IR absorbance spectra, $A(\nu)$, obtained from FDTD are compared with the IR absorbance spectra, $A_{\mathrm{M}}(\nu)$, obtained with a semi-analytical model developed by Davis et al. that was developed specifically to model IR microspectroscopy of a grating structure (periodic arrangement of slabs). [3] Importantly, this semi-analytical model implements the optical apparatus analytically and thus free of any numerical error.

The FDTD box is configured as follows. The FDTD box measures 200 µm in width (x and y) and 10 µm in height (z). The mesh cell size of the box is 100 nm, 24 PML layers are chosen at all six boundaries and a simulation time of 4000 fs is chosen. This configuration ensures accurate capturing of the scattered field from the grating structure (provided by the sufficient size of the box) and resolution of the molecular resonances present in the material (provided by the sufficient simulation time). A Gaussian beam of 0.6 NA is injected by the source S located $z = -1.6\ \mu\mathrm{m}$ in the forward (+z) direction. The start and stop wavelength of the source is 2.5 µm and 10 µm to cover molecular resonances of the sample in the mid-IR spectral range. To ensure an achromatic focus, 100 frequency dependent profiles are used. The grating structure consists of slabs of 30 µm width (x), infinite length (y) and 2 µm height, arranged in a period array using 30 µm spacing between the slabs. The width and height of the slab is chosen to mimic the size of typical substrate-adhered (flattened) cells. The slab material is modeled using the tabulated values for the refractive index of Matrigel, $n(\nu) + ik(\nu)$, a basement membrane matrix that is a representative model for IR molecular vibrations of cell samples (see Fig. 1 in the main text). Matrigel exhibits distinct and clearly identifiable molecular vibration modes, which makes it a suitable material for testing the FDTD method. The incident Gaussian beam interacts with the grating structure. The resulting scattered light is collected by a field domain profile monitor M located at $z = +1.6\ \mu\text{m}$. Spatial downsampling of a factor of 4 is used in the x,y directions to reduce the memory consumption of the simulation. The standard monitor settings are overridden by using a minimum sampling per cycle of 8, which reduces artifacts (eg. spikes) when sampling the scattered field. After the FDTD calculation has finished, light collection and refocusing is implemented using a near-field to far-field transform. The detector signal, $I(\nu)$, is obtained by spatial integration and the absorbance, $T(\nu)$, is calculated by normalization to a reference simulation, as explained in detail in the main text.

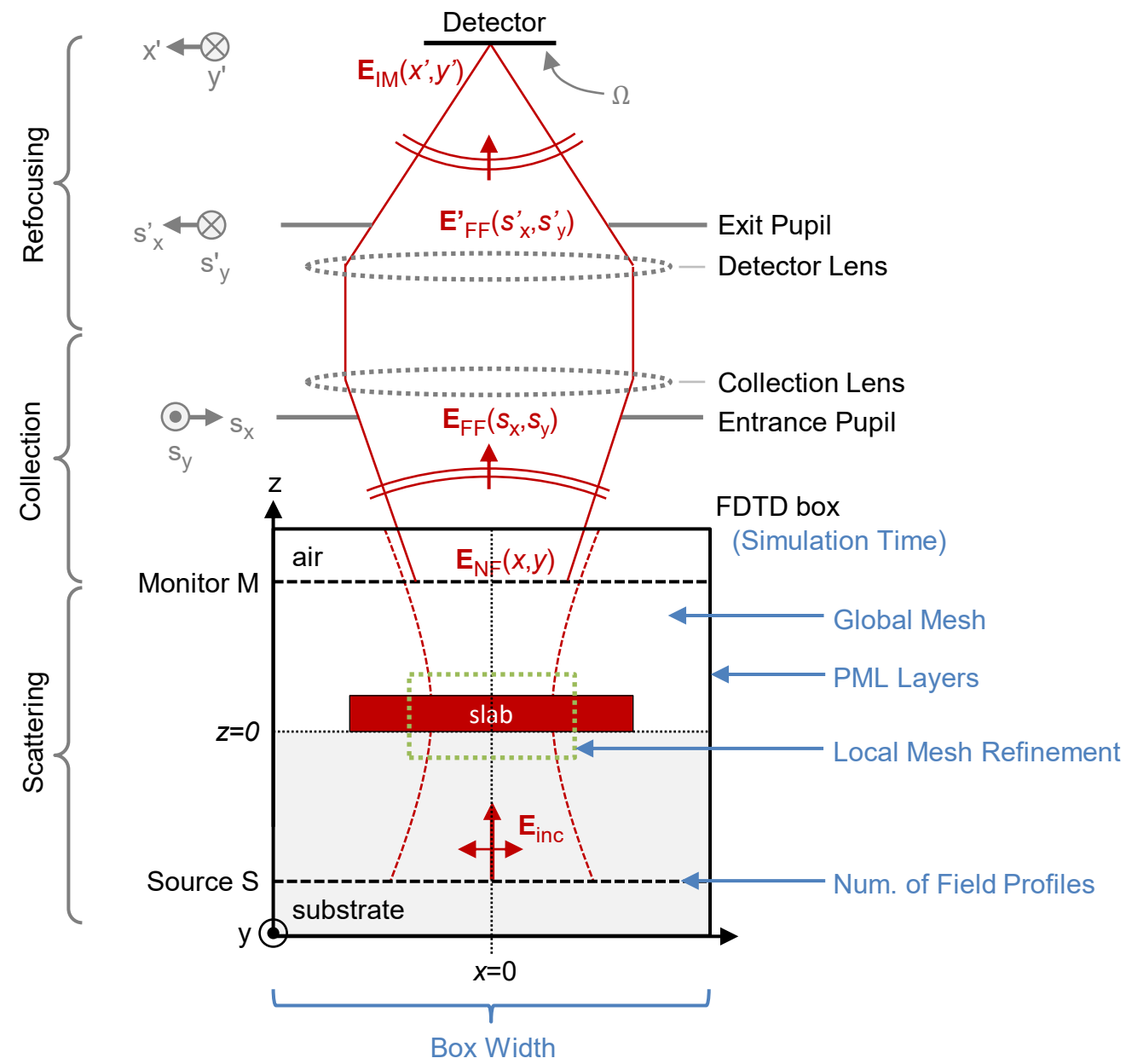

Figure S1. Computational IR microspectroscopy system as shown in Fig. 1 of the main text, but here in case of a grating structure consisting of a periodic array of flat homogeneous thin slabs. Additionally, the sources for numerical error are identified with the blue-colored labels.

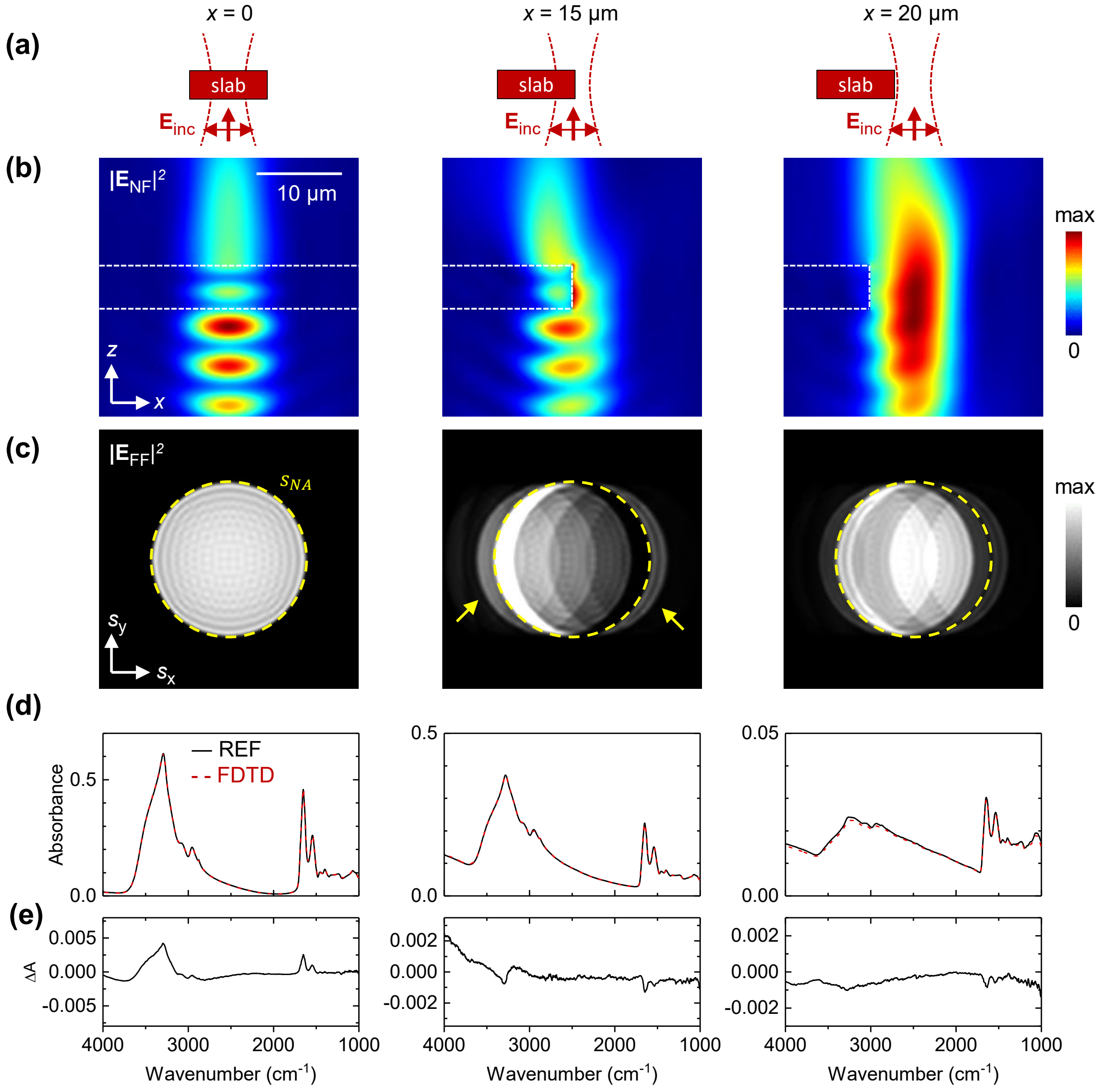

Figure S2: Illustration of the signal formation of the computational IR microspectroscopy method with the example of a grating structure that consists of a periodic array of homogeneous thin slabs made of Matrigel. (a) Various focus positions on the slab produce different amounts of light scattering, which serves to test the optical apparatus of the computational IR microspectroscopy method: $x = 0$ (left column), $x = 15\ \mu\mathrm{m}$ (middle column) and $x = 20\ \mu\mathrm{m}$ (right column). (b) Field distribution in the XZ plane (individually normalized), $\mathbf{E}^{\mathrm{NF}}(x, y, \nu)$. (c) Transmitted fields as obtained in the far zone, $\mathbf{E}^{\mathrm{FF}}(s_x, s_y, \nu)$ The dashed circle illustrates the collection cone of the objective as defined by its numerical aperture. Data plotted at $\nu = 1{,}200\ \mathrm{cm}^{-1}$ where the slab is only weakly absorbing. (d) Calculated IR absorbance spectra, $A(\nu)$ (red dashed line) as obtained from FDTD and comparison with the reference provided by the semi-analytical model, $A_{\mathrm{M}}(\nu)$ (black solid line). (e) Difference spectrum, $\Delta A(\nu) = A(\nu) - A_{\mathrm{M}}(\nu)$, revealing that the FDTD model from the semi-analytical model is less than 0.5% (peak difference in absorbance signal, $\Delta A(\nu)$) and of the order of 0.1% (root mean square error of $\Delta A(\nu)$).

***Illustration of signal formation with a grating structure***

Figure S2 illustrates the signal formation in case of modeling the grating structure using the computational IR microspectrometer method. In the following, three situations are considered, where the focus is placed at different locations on the slab to yield different amounts of light scattering (Fig. S2(a)). Figure S2(b) shows the corresponding electric field distribution, $\mathbf{E}^{\mathrm{NF}}(x, y, \nu)$, in the FDTD computational domain (a vertical cross section is shown). When the focus is positioned on the slab center ($x = 0$), a standing wave pattern can be observed inside the slab, which confirms the expected thin-film interference inside of the slab. The beam is partially transmitted through the sample (upper part of the FDTD box). It can be recognized that the beam does not suffer any major modification, indicating that light scattering is low. If, however, the focus is placed near or at the edge ($x = 15\ \mu\mathrm{m}$ and $x = 20\ \mu\mathrm{m}$), the structure of the transmitted beam is distorted significantly, indicating strong light scattering. The incident beam is also partially reflected at the lower interface of the slab, leading to a standing wave pattern in the bottom part of the FDTD box, which is not further investigated here. Figure S2(c) shows the associated fields at the far zone, $\mathbf{E}^{\mathrm{FF}}(s_x, s_y, \nu)$. It is immediately visible that the strong scattering for the cases $x = 15\ \mu\mathrm{m}$ and $x = 20\ \mu\mathrm{m}$ leads to light escaping the collection cone (fields outside the dashed circle in Fig. S2(c)). In comparison, the transmitted beam falls entirely inside the collection cone in the case of $x = 0$, confirming that light scattering is very low. Figure S2(d) shows the corresponding absorbance spectra, $A(\nu)$. As expected, the strong light scattering outside the collection cone produces a significant baseline in $A(\nu)$ in the cases of $x = 15\ \mu\mathrm{m}$ and $x = 20\ \mu\mathrm{m}$, while light scattering is only weak in case of $x = 0$ , yielding a near-zero baseline in $A(\nu)$. [3]

***Semi-analytical model provides a reference***

To validate the accuracy of the presented FDTD model, we compare with the results of a semi-analytical model from ref. [3]. In this semi-analytical, the grating structure is modeled using the coupled-wave theory, where the dielectric permittivity of the slab material is represented as a Fourier Series in the *x*-direction, corresponding to the direction of periodicity. The Fourier Series needs to be truncated to a finite number of terms $N$, where $N = 1000$ has been chosen for maximum numerical accuracy within the bounds of feasible computation. The optical apparatus was configured identically to the FDTD method. The source code of this method was obtained from the corresponding author from ref. [2] and the code was run on Matlab. Figure S2(d) shows that the presented FDTD model, $A(\nu)$ (red line), reproduces accurately the absorbance spectra obtained with the semi-analytical model, $A_{\mathrm{M}}(\nu)$ (black line). Figure S2(e) shows the difference spectra, $\Delta A(\nu) = A(\nu) - A_{\mathrm{M}}(\nu)$, illustrate that the deviation is less than 0.5% (peak value) and of the order of 0.1% (root mean square error of $\Delta A(\nu)$).

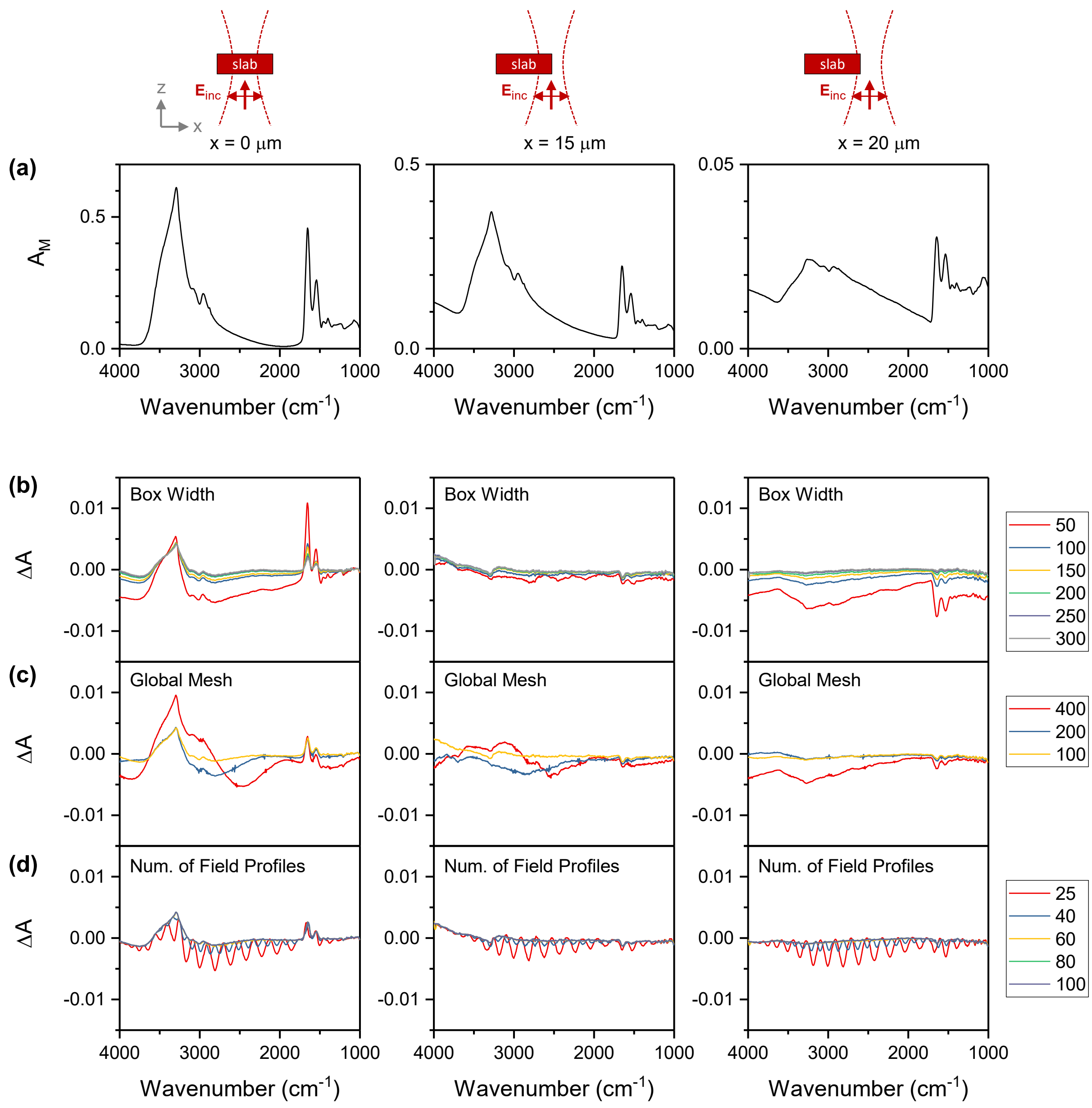

Figure S3. Convergence study of the computational IR microspectroscopy system using a grating structure (slabs) as a test sample. (a) Absorbance spectra, $A_{\mathrm{M}}(\nu)$, obtained with the semi-analytical model for different displacements of the slab, $x$, with respect to the optical axis. (b) Error spectra, $\Delta A(\nu)$, for a variation of the width of the FDTD box (values in the legend are in $\mu$m). (c) Error spectra, $\Delta A(\nu)$, for a variation of the global mesh size (values in the legend are in nm). (d) Error spectra, $\Delta A(\nu)$, for a variation of the number of field profiles of the source, which determines the achromaticity of the Gaussian focus.

***Convergence Test***

Achieving good convergence of the FDTD simulations requires the setting of several FDTD-specific parameters. In the following, we provide a detailed convergence study and show how to identify sources for numerical error and evaluate their significance.

As before in Fig. S2, we consider the three cases where the beam is focused at the center of the slab ($x = 0$), on the edge ($x = 15\ \mu\mathrm{m}$) and away from the edge ($x = 20\ \mu\mathrm{m}$). Figure S3(a) shows the absorbance spectra, $A_\mathrm{M}(\nu)$, obtained with the semi-analytical model for reference. We first focus on the following sources for numerical error: the width of the FDTD box, the global mesh size and the number of field profiles of the source are the main sources for error in case of the computational IR microspectroscopy method (see blue labels in Fig. S1). To evaluate the errors of these sources, we calculate the absorbance, $A(\nu)$, for a variation of the selected parameter and compare it to the model absorbance, $A_\mathrm{M}(\nu)$. The error spectrum is then given as the difference between both quantities,

$$\Delta A(\nu) = A(\nu) - A_\mathrm{M}(\nu)\ , \qquad (1)$$

Figure S3(b) shows the error spectra, $\Delta A(\nu)$ (Eq. (1)), for a variation of the width of the FDTD box. Small box widths (e.g. $W = 50\ \mu\mathrm{m}$, red spectra) lead to a large difference in absorbance of the order of 1%. This error can be reduced significantly by choosing larger box widths. This behavior can be attributed to the fact that the Gaussian focus is implemented by a 2D planar source S and light collection is implemented by a 2D planar monitor M. A large FDTD box is needed to (i) avoid clipping of any intensity of beam profile at the source S, which otherwise would introduce beam scattering on injection, (ii) accurately collect the obliquely scattered fields with monitor M.

Figure S3(c) shows the error spectra, $\Delta A(\nu)$ (Eq. (1)), for a variation of the global mesh size. The coarse global mesh setting of 400 nm (red spectra) lead to significant deviations of up to 1%. This behavior is attributed to numerical dispersion that is introduced by the coarse mesh and becomes more noticeable for higher wavenumbers. As a rule of thumb, it is widely considered that there should be at least 10 mesh cells per wavelength. In case of the shortest wavelength considered here (4000 $\mathrm{cm}^{-1}$, $\lambda = 2.5\ \mu\mathrm{m}$), it is thus recommended to use a maximum mesh size of $\Delta x = \lambda/10 = 250\ \mathrm{nm}$ or better.

Figure S3(d) shows the error spectra, $\Delta A(\nu)$ (Eq. (1)), for a variation of the number of field profiles of the source S. This parameter refers to a specific feature of the commercial FDTD package, which contemplates frequency dependent profiles for sources to provide achromatic focusing of Gaussian beams, that is, each frequency component is focused to the same plane, $z = 0$. It is apparent from the spectra, that the low number of 25 field profiles is not sufficient to provide achromatic focusing across the considered spectral range from 1000 to 4000 $\mathrm{cm}^{-1}$, resulting in oscillator deviations in the error spectra, $\Delta A(\nu)$.

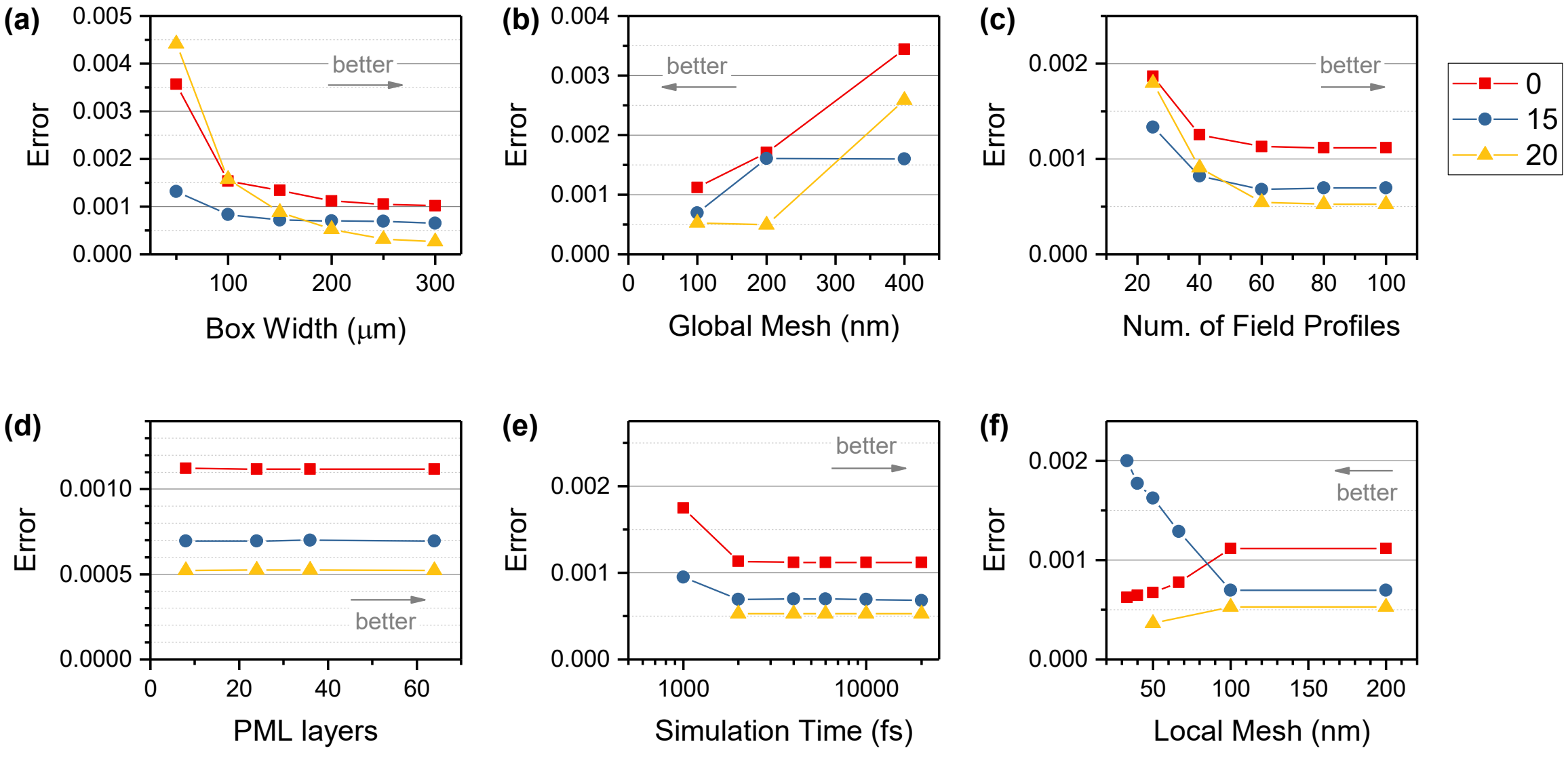

Figure S4. Summary of the convergence study of the computational IR microspectroscopy system, showing the root mean square error, $A_{\mathrm{err}}$, for each parameter variation: (a) FDTD box width, (b) global mesh size, (c) number of field profiles of the source, (d) number of PML layers, (e) simulation time and (f) local mesh size. Panels (a-c) are calculated from the data shown in Fig. S3(b-d) and reflect the most prominent sources for errors. Panels (d-e) reflect further, minor sources of error.

Having illustrated the error spectra, $\Delta A(\nu)$, in Fig. S3, we now proceed to quantify the error. To this end, we calculate the absolute error as the root mean square of the difference between both spectra,

$$A_{\mathrm{err}} = \sqrt{\sum_{n=1}^{N} (A(\nu_n) - A_{\mathrm{M}}(\nu_n))^2}\,, \qquad (2)$$

where the summation is performed over the N discrete frequency points of the spectra $A(\nu_n)$ and $A_{\mathrm{M}}(\nu_n)$. The parameter variation is conducted such that initial simulation considers the best set of parameters and then the selected parameter is subsequently tuned to produce worse results.

Figure S4 shows that the main sources of error are the FDTD box width, global mesh settings and number of field profiles of the source, which show the largest variation in $A_{\mathrm{err}}$ (Fig. S4(a-c)). It is apparent that the root mean square error, $A_{\mathrm{err}}$ (Eq. (2)), can be reduced to below $0.002$ by choosing a sufficiently large FDTD box width (at least 100 $\mu$m), using a global mesh size of 200 nm (or better) and using at least 60 field profiles for the source. Figure S4(d-f) shows further sources for errors that are typically considered in FDTD convergence studies. We find that the minimum number of PML layers of 8 already yields low error and further increasement is not warranted (Fig. S4(d)). Further, a simulation time of 2000 fs is sufficient to model the molecular vibrations of matrigel and thus most biological matter (Fig. S4(e)). Optionally, it could be desirable to locally define a finer mesh setting around the sample in order to better model fine topographical features (this local mesh refinement is indicated by the gray dashed box in Fig. S1(a)). In case of the slab considered here, a finer mesh can help to decrease the error in general (Fig. S4(f)). The error was made worse (increased) in the case of focusing on the edge of the slab ($x = 15\ \mu\mathrm{m}$) for unknown reasons.

In conclusion, if properly configured, the computational IR microspectroscopy method can deliver accurate predictions of IR absorbance spectra. The convergence study presented above details how to identify source of numerical error and to determine the best settings for best accuracy while maintaining reasonable computation effort.